\documentclass[10pt,twocolumn,twoside]{article}
\usepackage{graphics,array,afterpage}
\newcommand{\etal}{ {\it et al.} }
\newcommand{\cthead}[1]{\multicolumn{1}{c|}{#1}}
\renewcommand{\section}[1]{\par\noindent\parbox{\linewidth}{\centering #1}}
\newcommand{\lb}{\linebreak}
\newcommand{\nlb}{\nolinebreak}

\newcommand{\tablecaption}{\hline
\cthead{\rule{0mm}{5mm}Transition}&Frequency,\rule{-3mm}{0mm} &
\cthead{1}& &\multicolumn{1}{c||}{2}&
\cthead{Transition}&Frequency,\rule{-3mm}{0mm} &
\cthead{1}& &\multicolumn{1}{c}{2} \\
&\cthead{\rule[-3mm]{0mm}{5mm}GHz}&\cthead{$\tau$}& &\multicolumn{1}{c||}{$\tau$}&
&\cthead{GHz}&\cthead{$\tau$}& &\multicolumn{1}{c}{$\tau$}\\
\hline
}

\makeatletter
\renewcommand{\@biblabel}[1]{}
\makeatother

\makeatletter
\renewcommand{\@oddhead}{\hfil A MODIFICATION OF THE MONTE
CARLO METHOD FOR SIMULATION\hfil\thepage}
\renewcommand{\@oddfoot}{ASTRONOMY LETTERS~~~~Vol. 25~~~~~No.~3~~~~~
1999\hfil}
\renewcommand{\@evenhead}{\thepage\hfil VORONKOV\hfil}
\renewcommand{\@evenfoot}{\hfil ASTRONOMY LETTERS~~~~Vol. 25~~~~~No.~3~~~~~
1999}
\def\tiny{\@setsize\tiny{6pt}\vipt\@vipt}
\makeatother

\begin{document}
\thispagestyle{empty}
\twocolumn[
\centering
\parbox{\linewidth}{\raisebox{0.6cm}[0mm][0mm]{
\parbox{\linewidth}{\tiny\it Astronomy Letters, Vol. 25, No. 3, 1999,
pp. 149-155. Translated from Pis'ma v Astronomicheskii Zhurnal, Vol. 25,
No. 3, 1999, pp. 186-193.\\
Original Russian Text Copyright \copyright 1999 by Voronkov.
\vskip 3mm
\hrule
\vskip 1mm
\hrule
}}
\raisebox{-22.8cm}[0mm][0mm]{
\parbox{\linewidth}{\hfil 1063-7737/99/2503-0149\$15.00~\copyright~
MAIK "Nauka/Interperiodica"\hfil}}
}
\parbox{0.9\linewidth}{
\centering

{\Large\bf
A Modification of the Monte Carlo Method for Simulation\linebreak of Radiative
Transfer in Molecular Clouds

\vskip 5mm
\large
M.A.Voronkov$\bf^{1, 2}$

}
\vskip 5mm
$^1$ {\it Moscow State University, Moscow, 119899 Russia}\\
$^2$ {\it Astrospace Center, Lebedev Physical Institute, Russian Academy of
Sciences,\linebreak ul. Profsouznaya 84/32, Moscow, 117810 Russia}\\
\vskip 5mm
Received August 24, 1998
\vskip 5mm
\centering
\parbox{\linewidth}{
\small
{\bf Abstract}--We propose a method of simulation that is based on the 
averaging of formal solutions of the transfer equation by taking the integral
by the Monte Carlo method. This method is used to compute two models,
\nlb which 
correspond to the limiting cases of hot gas and cold background 
radiation and of hot background radiation~and\lb cold gas, for E-methanol 
emission from a compact homogeneous spherical region. We analyse model 
level pop\-ulations by using rotational diagrams in the limiting cases 
mentioned above. Model optical depths of the lines with frequencies below
300 GHz up to $J$=11 inclusive are given.
}}
\vskip 5mm
]

\section{INTRODUCTION}

Molecular methanol (CH$_3$OH) emission, along with emission of other
molecules, is commonly observed toward\- star-forming regions. 
Some molecular methanol transitions produce narrow and intense 
maser lines in some sources, while, in other sources, these transitions 
produce less intense or no detectable lines; at the same time, 
maser activity may show up in other transitions. The conditions for the 
formation of methanol masers can be studied by numerically simulating
radiative transfer. In such problems, we simultaneously solve the 
transfer\- equation and the system of statistical-equilib\-rium equations\- 
for a large number of lines. This problem is commonly solved either 
by the Monte Carlo\lb (MC) method or by large velocity gradient (LVG)
method. In contrast to the LVG method, the MC\lb method allows computations
to be performed for a compact\- cloud and a small velocity gradient.
In addition, the model adopted in the MC method admits several 
important complications: inhomogeneity and complex structure of an 
emitting cloud and a more realistic\- allowance\- for the effect of its 
nonspherical geometry (with no use of factor $\varepsilon^{-1}$).
However, the MC\lb method requires more computing time than the LVG 
method does. The MC method, whose algorithm was described by Bernes (1979),
is based on the replacement of the real radiation field by a number of
model\lb photons and on the simulation of their propagation through the 
medium with the computation of the number of molecular excitations in
each of the shells into which the cloud is divided. The classical MC
method assumes that model photons are emitted uniformly in\lb\lb all
directions in each of the shells and come from the outside. Juvela (1997) 
proposed a modification of the standard MC method: the paths of all
model photons begin at the cloud edge. However, as in the classical\lb
MC method, the formalism of model photons is used as the basis for the 
computations. In this paper, we make an attempt to deduce the method 
from the transfer\lb equation while keeping the main features of the MC
method and propose a method that is based on the aver\-aging of formal 
solutions to the transfer equation by taking the integral by 
the Monte Carlo method. This method was tested on simulations of E-methanol
emission from a compact ($\sim$0.005~pc) spherical cloud with\lb no
velocity gradient.

\vskip 2mm
\section{THE ALGORITHM AND BASIC APPROXIMATIONS}
\enlargethispage{-\baselineskip}
\vskip 2mm

We performed simulations in the simplest case of a static spherical cloud 
with a uniform distribution of the H$_2$ and CH$_3$OH number densities and
kinetic temperature. The influence of dust was ignored. The spectrum of
the background radiation was assumed to follow the Plank law. The dilution
of the background radiation was disregarded. The radius of the cloud 
was set equal to 1.5$\times 10^{16}$~cm.

In the algorithm under consideration, as in the~algo\-rithm of Bernes (1979),
the solution of the problem reduces\- to an iterative procedure which is
implemented using the system of statistical-equilibrium equations\lb (\ref{stateq})
and which yields new approximations for the popu\-lations

\begin{equation}
\label{stateq}
\left\{
\begin{array}{l}
\sum\limits_kn_k\{B_{kj}\overline{I}+A_{kj}\delta_{kj}+C_{kj}\}\\
=n_j\sum\limits_k\{A_{jk}\delta_{jk}+B_{jk}\overline{I}+C_{jk}\}\\
\sum\limits_kn_k=const\mbox{,}\\
\end{array}
\right.
\end{equation}
where $\delta_{kj}=1$ when and only when the relation between the energy
levels $E_k>E_j$ holds; otherwise, $\delta_{kj}=0$; $A_{jk}$ and $B_{jk}$
are the Einstein coefficients (which are assumed to be zero for forbidden
transitions); $C_{jk}$ are the colli\-sional constants; $n_k$ is the 
population of the $k$th level; and $\overline{I}$ is the average 
intensity. In the first equation in\lb (\ref{stateq}), the subscript $j$ changes 
in the range $1<j<(N-1)$, where $N$ is the total number of levels involved.
The average intensity is given by
\begin{equation}
\label{averI}
\overline{I}=\frac12\int\limits^{+\infty}_0f(\nu)\;d\nu\int\limits_{-1}^{+1}
I(\nu,\mu)\;d\mu\mbox{,}
\end{equation}
where $\nu$ is the frequency, $\mu$ is the cosine of the angle\lb
between the radius and the selected direction, and $f(\nu)$ is the 
line profile (here it was assumed to be the Dop\-pler profile).
This integral can be calculated by the Monte~Carlo method by
simulating the random vari\-ables~$\nu$ and $\mu$ with Gaussian
(for the Doppler line pro\-file)~and uniform distributions, respectively:
\begin{equation}
\label{nu}
\frac{\nu}{\nu_0}-1=\frac 1c\left(\sigma\sin\left(2\pi R\right)
\left(-\ln R'\right)^{1/2}+v(r)\mu\right)
\end{equation}
\begin{equation}
\label{mu}
\mu=1-2R''\mbox{,}
\end{equation}
where $R$, $R'$ and $R''$ are independent random variables with a
uniform distribution in the interval 0 to 1; $\nu_0$ is\lb the rest
frequency of the transition under consideration; $v(r)$ is the velocity
field at a given point; and $\sigma$ is the Doppler halfwidth.
The average intensity can be\lb replaced by the mean
\begin{equation}
\label{MatW}
\overline{I}=MI\left(\mu,\nu\right)\mbox{.}
\end{equation} 
The averaging over sufficiently thin shells into which\lb the cloud 
is divided leads to an additional integration~of
(\ref{averI}) over the radius, which reduces to the simulation of yet
another random variable that gives the distance\lb from the cloud center
to the common point of the bundle of directions used to compute the mean
(\ref{MatW}):
\begin{equation}
\label{Rdis}
r=\left\{r_{in}^3+R(r_{out}^3-R_{in}^3)\right\}^{1/3}\mbox{,}
\end{equation}
where $r_{in}$ and $r_{out}$ are the radii of the inner and outer
boundaries of the shell, and $R$ is a random variable that\lb is uniformly
distributed in the interval 0 to 1. The inten\-sity $I\left(\mu,\nu\right)$
for a given direction can be estimated by using

\begin{equation}
\label{IntEq}
\begin{array}{c}
I_\nu\left(\mu\right)=I_{\nu\;bg}\left(\mu\right)\exp\left\{-\int\limits_0^L
x_\nu(y)\;dy\right\}\\
+\int\limits_0^L\varepsilon_\nu(X)\exp\left\{-\int
\limits_0^Xx_\nu(y)\;dy\right\}\;dX\mbox{,}\\
\end{array}
\end{equation}
where $x_\nu$ and $\varepsilon_\nu$ are the absorption and emission
coefficients of the medium, respectively; $L$ is the distance to the
cloud edge in the direction of integration; and $I_{\nu\;bg}$~is the
intensity of the background emission for a given\lb direction. The
integration path is broken up into segments in which the emission
and absorption coeffi\-cients~can be assumed to be constant (in the
presence of\lb a velocity gradient, this path can be much shorter than
the path within a single shell where the populations are assumed
to be constant). The change in intensity within a segment can be obtained
by integrating (\ref{IntEq}):
\begin{equation}
\label{DI}
\Delta I_\nu=\left(\frac{\varepsilon_\nu}{x_\nu}-I_{in}\right)\left(
1-e^{-x_\nu l}\right)\mbox{,}
\end{equation}
where $I_{in}$ is the intensity at the beginning of the segment, and
$l$ is the segment length in the direction~under\- consideration. The
emission and absorption coeffi\-cients~are~expressed\- in terms of the
level populations
\begin{equation}
\label{Eps}
\varepsilon_\nu=\frac{h\nu}{4\pi}f(\nu)A_{ul}n_u\mbox{,}
\end{equation}
\begin{equation}
\label{Kappa}
x_\nu=\frac{c^2A_{ul}f(\nu)}{8\pi\nu^2}\left\{\frac{g_u}{g_l}n_l-n_u\right\}\mbox{,}
\end{equation}
where $A_{ul}$ is the Einstein coefficient for the corresponding
transition; $n_u$, $g_u$ and $n_l$, $g_l$ are the populations~and
statistical weights of the upper and lower levels, respectively. Thus,
the average intensity is expressed in terms of the level populations
and the intensity of the background radiation, which allows us to
implement the\lb iterative process using the system of
statistical-equilib\-rium~equations (\ref{stateq}) for the populations.
When the itera\-tive process has converged, the derived populations are used
to compute the intensity of the cloud radiation (or brightness temperature)
with the aid of the formal solution to the transfer equation~(\ref{IntEq}).

\vskip 5mm
\section{RESULTS}
\vskip 2mm

We performed the computations for two models with kinetic temperatures
of 70 and 20 K and with background-radiation temperatures of 2.7 and 70 K,\lb
respectively (below referred to as models I and II\lb respectively). The
methanol column densities in~the\lb two models were assumed to be the same
and~equal~to\lb $1.5\times10^{15}$~cm$^{-2}$ (the radius is $1.5\times10^{16}$~cm,~
the~H$_2$~num\-ber density is $10^5$~cm$^{-3}$, and the methanol~abundance~is\nopagebreak\lb
$10^{-6}$). When computing the models,~we~took~into\nopagebreak\lb account 124 lower
rotational levels of the ground~tor\-sional state of E-methanol
($J<15$, $|K|<6$, and $E<200$~cm$^{-1}$).

Torsional transitions were ignored. We computed\lb the energies of the
levels and Einstein coefficients\lb ($A$) by using approximate formulas from
Pikett\etal (1981). The collisional constants are known poorly and
were computed here in the gas-dynamical approxima\-tion~of Lees (1974).
The iterations were terminated at an accuracy of 0.1\% of the population.

\vskip 5mm
\section{DISCUSSION}
\vskip 2mm
\hyphenation{according formula different substituting relations}

The algorithm of Bernes (1979) uses the quantity $S_{lu,m'}$ (for the
$m'$ shell and for the transition between~$l$ and $u$) which is proportional
to the intensity instead of the latter; this quantity is accumulated
over all model photons in all shells. A comparison of the system of\lb
statistical-equilibrium equations yields~the~relation\lb
$\left(\sum S_{lu,m'}\right)\!\!=\!\!\overline{I}_{lu,m'}B_{lu}$.~In~each~step,~
according~to~formula\lb (6) from Bernes (1979), the following quantity is
added\lb to $\left(\sum S_{lu,m'}\right)$:
\begin{equation}
\label{BernesS}
\begin{array}{c}
S_{lu,m'}\\\\
=\frac{h\nu}{4\pi}f(\nu)B_{lu}\frac{s_kW_0\exp\left(-\sum\limits_{i=1}
^{k-1}\tau_i\right)}{V_{m'}\tau_k}\left\{1-\exp\left(-\tau_k\right)\right\}
\makebox[-0.5mm]{,}\\
\end{array}
\end{equation}
where $W_0$ is the initial weight of the model photon; $V_{m'}$ is the
volume of the $m'$ shell; and $\tau_i$ and $s_i$ are the optical\- depth
and path of the model photon in the $i$th step,\lb respectively. In order
to compare the method of Bernes (1979) with the method proposed here,
which we\lb deduced  from the transfer equation, let us consider the
contribution of the intrinsic emission from the $m$ shell,\lb
the emission from a different~$m'$~shell,~and~the~background radiation
to $\left(\sum S_{lu,m'}\right)$.~Substituting~the~relations\lb for $\tau_1$ and
for the initial weight of the model photon\lb produced in the $m'$ shell,
$W_0\!\!=\!\!V_{m'}A_{ul}n_u/N_{ph}$,~where~$N_{ph}$ is the total number of
model photons, we obtain the~fol\-lowing expression for the contribution of the
intrinsic emission from the $m'$ shell:
\begin{equation}
\label{SMpRad}
S_{lu,m'}=\frac {B_{lu}}{N_{ph}}\left(\frac{2h\nu^3}{c^2}\frac 1{
\frac{n_lg_u}{n_ug_l}-1}\right)\left\{1-\exp\left(-\tau_1\right)\right\}\mbox{.}
\end{equation}
Since the term in parentheses is the source function and comparing
this equation with (\ref{DI}), we conclude that the contribution of the
intrinsic emission is accurately represented in the method of
Bernes (1979). It follows\lb from (\ref{IntEq}) that the contribution
of the background radiation must be given by
\begin{equation}
\label{SExtRad}
S_{lu,m'}=\frac {B_{lu}}{N_{ph}}I_{bg}\exp\left(-\sum\limits_{i=1}
^{k-1}\tau_i\right)\mbox{,}
\end{equation}
where $I_{bg}$ is the intensity of the background radiation~at the
frequency of the transition between levels $u$ and $l$. If the cloud
is divided into shells of equal volume ($V_{m'}=const$), then the
path of the model photon is the same in each step ($s_k=const$) and
small enough for the optical depth in each step to be small in
absolute~value.~For\-mula~(\ref{SExtRad}) can then be derived from
(\ref{BernesS}) by using the linear approximation for the exponent.
The~contribu\-tion~of~the~emission from a different shell can be
obtained from (\ref{BernesS}), by analogy with (\ref{SMpRad}),
\begin{equation}
\label{SMRad}
\begin{array}{c}
S_{lu,m'}=\frac {B_{lu}}{N_{ph}}\frac{V_m}{V_{m'}}\exp\left(-\sum\limits_{i=1}
^{k-1}\tau_i\right)\\
\times\left(\frac{2h\nu^3}{c^2}
\frac 1{\frac{n_lg_u}{n_ug_l}-1}\right)_k\left\{1-\exp\left(-\tau_k\right)\right\}\mbox{,}
\end{array}
\end{equation}
where the subscript $k$ of the source function means that this function
is calculated from the populations in the $k$th step
(in the $m'$ shell rather than the $m$ shell where the model photon was
formed). It follows from (\ref{IntEq}) that this contribution must be
equal to the product of (\ref{SMpRad}), where the source function is
calculated from the populations in the first step and the exponential
factor from (\ref{SExtRad}). If the cloud is divided into shells of equal
volume and if the optical depth in each step is small in absolute
value (which allows a linear expansion of the expo\-nent),~then, using
relation (\ref{Eps}) for the emission coeffi\-cient,~we~obtain the last
condition for the applicability of the algorithm of Bernes (1979):
\begin{equation}
\label{Cond3}
f_k(\nu)(n_u)_ks_k\approx f_1(\nu)(n_u)_1s_1\mbox{,}
\end{equation}
where the subscripts denote the first and $k$th steps in the
propagation of the model photon. Relation (\ref{Cond3}) is the condition
for the equality of the specific column densi\-ties~of~the emitting molecule
in each step in the~direc\-tion~of~propagation of the model photon.
The latter~con\-dition is most stringent and is difficult to satisfy in
practice. Since uncertainty in the estimate of the radiation field
in the algorithm of Bernes (1979) produces additional noise in the method,
the use of the algorithm outlined here seems more appropriate.
Our method correctly describes the radiation field (the formulas follow
from the transfer equation); therefore, its application is restricted
only by the convergence of the iterative pro\-cedure used (matches the
iterative procedure in the\lb classical MC method) and by the available
computing time. In practice, the convergence can be hampered in the
presence of strong masers.
\begin{figure}[t]
\resizebox{\hsize}{!}{\includegraphics{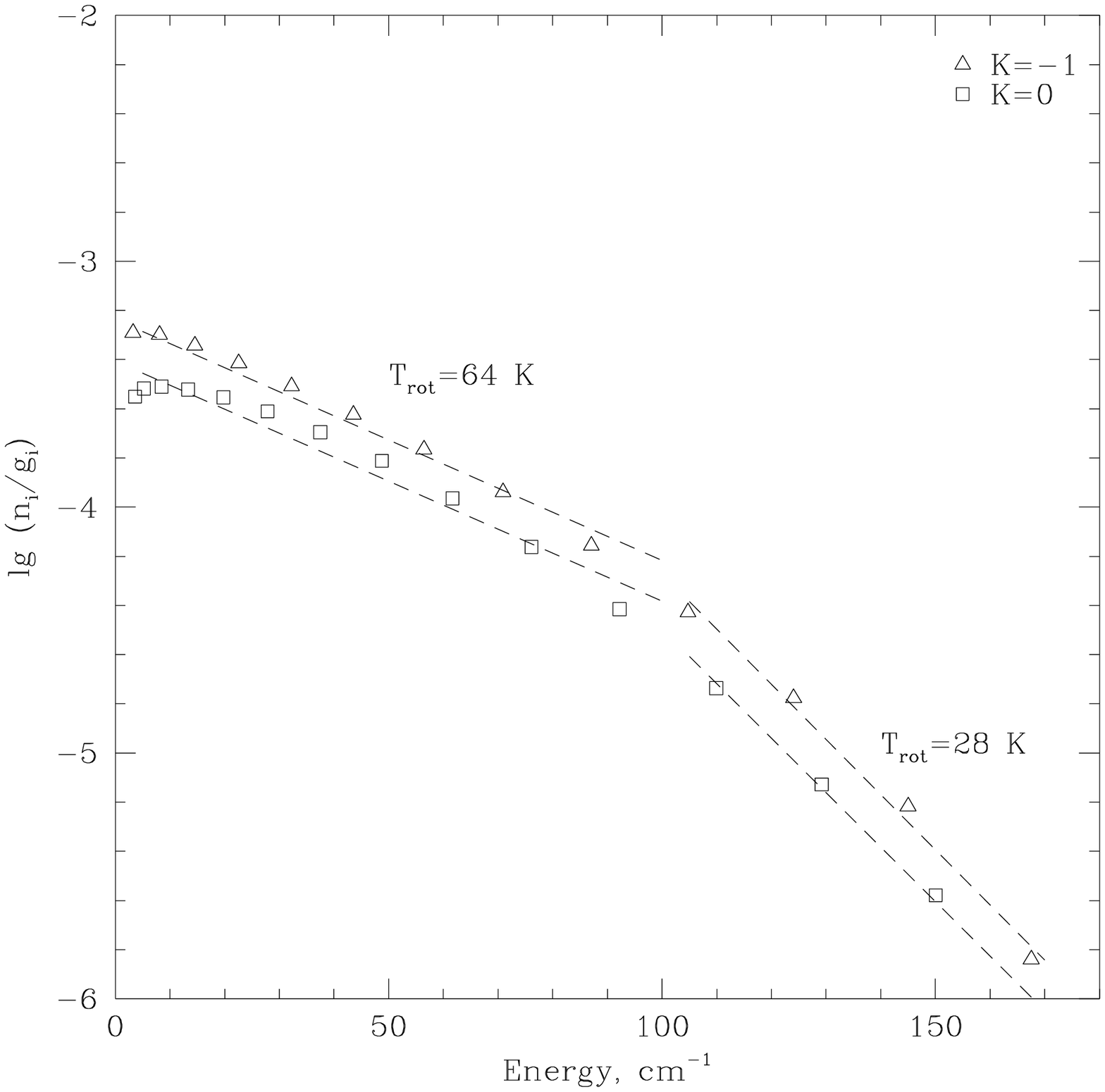}}
\centering
\addtocounter{figure}{1}
\vskip -5mm
\parbox[t][1.31cm]{0.93\hsize}{\footnotesize{\bf Fig. \thefigure.} Logarithm of the ratio of the
population~(in~cm$^{-3}$)~to the statistical weight versus the energy of
a given level for\lb the model with a background-radiation temperature of 2.7~K
and a cloud kinetic temperature of 70~K. The dashed lines correspond to
rotational temperatures of 64 and 28~K.}
\vskip -5mm
\end{figure}
\begin{figure}[t]
\resizebox{\hsize}{!}{\includegraphics{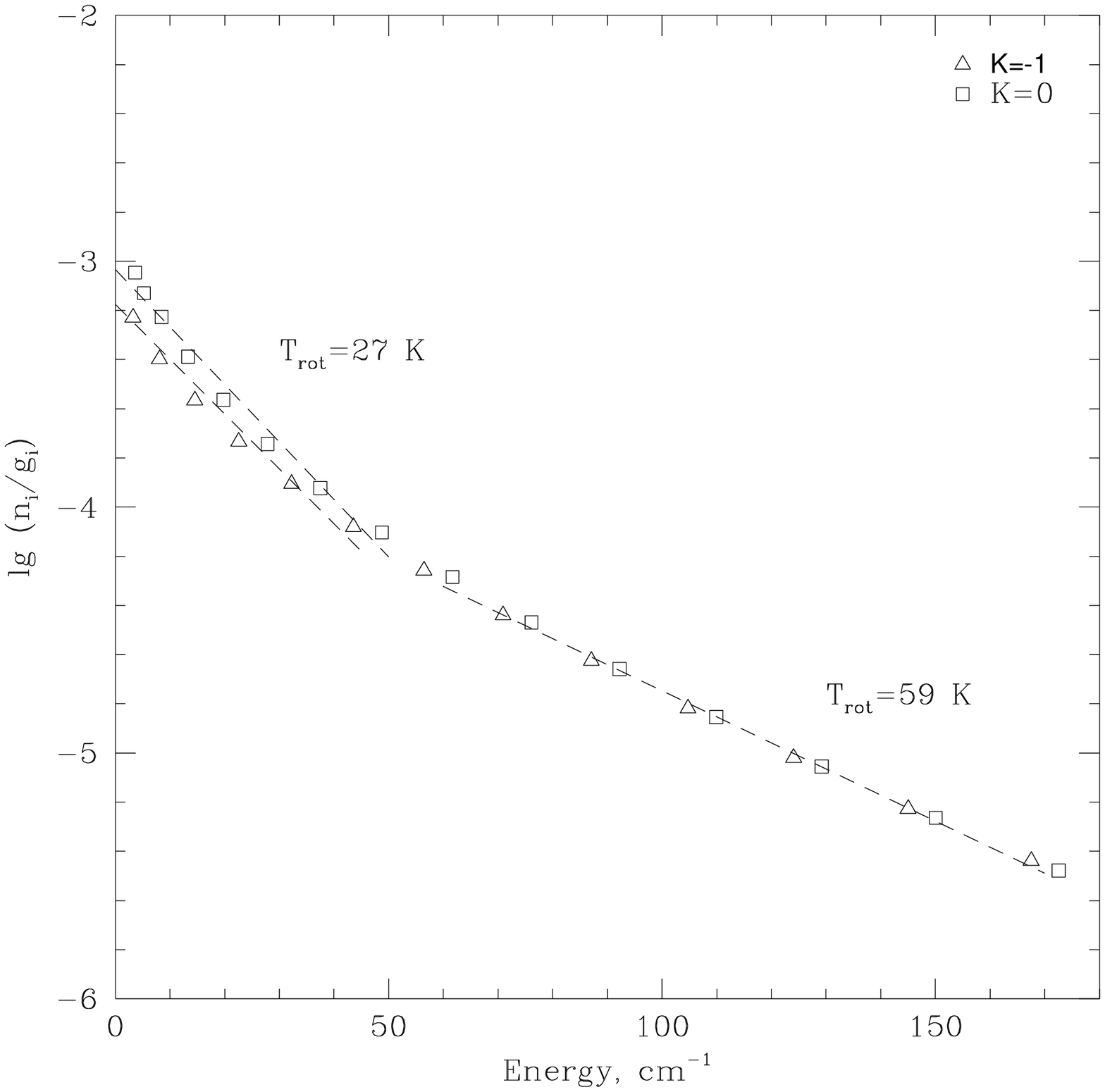}}
\centering
\addtocounter{figure}{1}
\vskip -5mm
\parbox[t][1.31cm]{0.93\hsize}{\vfil\footnotesize{\bf Fig. \thefigure.} Same as Fig.~1 for
the model with a background-radi\-ation temperature of 70~K and a cloud
kinetic temperature\lb of 20~K. The dashed lines correspond to rotational temperatures
of 27 and 59~K.}
\vskip -5mm
\end{figure}
\hyphenation{includes ladder series}

The qualitative behavior of populations in the limiting~cases of hot
gas (model I) and hot background radi\-ation (model II) is convenient
to analyze as follows. Let us consider the relation between the ratio
of the level population to its statistical weight and the energy of this\nopagebreak\lb
level on a logarithmic scale. In order not to overload~the\nopagebreak\lb picture,
let us consider only the ladders with quan\-tum~numbers $K=0$ and -1.
In the LTE case, the level population would be given by Boltzmann
formula,\lb and we would have a straight line whose slope would be
related to the temperature. In the non-LTE case, however, we have
a curve (Figs.~1 and 2). The levels that correspond to the
different ladders in these figures are indicated by different symbols.
In the model with hot gas, the $K=-1$ ladder is heavily overpopulated
relative to the $K=0$ ladder at small quantum numbers $J$, while in the
radiation-dominated model, the $K=0$ ladder lies above the $K=-1$ ladder;
i.e., there is an inversion of other transitions. This may correspond to
the division\lb of methanol masers into class I (Fig.~1) and
class II (Fig.~2) masers (Batrla\etal 1987; Menten\etal 1991).
The slope of the curve at energies $E<100$~cm$^{-1}$ in Fig.~1
corresponds to a rotational temperature $T_{rot}\sim64$~K,\lb which is
close to the kinetic temperature of the\lb medium. At high energies,
$T_{rot}\sim28$~K. In Fig.~2, rotational temperatures
$T_{rot}\sim27$~K (which is close to the kinetic temperature
of the medium) and $T_{rot}\sim59$~K (which is close to the
background-radiation tempera\-ture~$T_{bg}=70$~K) correspond to low
and high energies, respectively. Thus, the slope of the curve appears
to be mainly determined by collisions at low energies and by radiation at
high energies. The distribution of populations within the ladder (in each
of the segments) is close to the Boltzmann distribution, while, between
the ladders, it is an purely non-LTE distribution. For this reason,
no emission arises in a-type transitions ($\Delta K=0$).\lb The b-type
transitions (which occur with a change in\lb the quantum number K) tend
to produce a series of the form (J$+\alpha$)$_{K\pm1}-$J$_K$,
where $\alpha=0,\pm1$ (Menten\etal\lb 1986b). Since the series~
J$_0-$(J+1)$_{-1}$~includes~transi\-tions~from the $K\!\!=\!0$~ladder~to~
the~$K\!\!=\!\!-1$~ladder~and~vice\lb versa, some of the transitions in this series
exhibit  activity in model I ($4_{-1}-3_0$, $5_{-1}-4_0$, etc.), and
other are\lb active in model II ($0_0-1_{-1}$, $1_0-2_{-1}$, $2_0-3_{-1}$).
Masers~in\lb this series were detected in the observations of Turner\lb\etal
(1972), Zuckerman\etal (1972), Batrla\etal (1987), and
Slysh\etal (1997, 1999). In the models\lb under consideration, the series
J$_0-$J$_{-1}$ at 157~GHz gives a weak inversion of transitions for J=2
and 3. Masers\lb in this series were detected in the observations~for~
$2\!\!\le $J$\le\!\! 8$ (Slysh\etal 1995).
The transitions in the series
J$_0-$\lb(J-1)$_{-1}$ have higher frequencies and show no inversion in the
models under consideration. A similar behavior\lb of the populations is
observed for transitions between other ladders. The (J+1)$_0-$J$_1$
transitions exhibit class I and II activity for $J>2$ and $J<3$,
respectively.
Wilson\lb\etal (1985) detected a maser in the $2_1-3_0$ line
at 20~GHz. Slysh\etal (1992) identified weak thermal emission in the
$4_0-3_1$ line at 28~GHz. The series~J$_1-$J$_0$\lb at 165~GHz and
J$_1-$(J-1)$_0$ give no inversion in the\lb models under discussion. No
masers in the series J$_1-$J$_0$ were detected in the observations
of Slysh\etal (1999). The transitions in the series J$_2-$J$_1$ at
25~GHz give an\lb inversion in model I. No masers in this series were\lb
detected in the observations of \hfil Barret\etal (1971) \hfil and
\onecolumn

\vskip 3mm
\parbox[t][\textheight]{\linewidth}{
\hyphenation{kinetic abundances densities number}
Peak optical depths of the lines for models with the following H$_2$ number
densities, methanol abundances, kinetic temperatures, and
background-radiation temperatures: 1 - $n_{H_2} = 10^5$~cm$^{-3}$,
$X=10^6$, $T_{kin}=70$~K, and $T_{bg}=2.7~K$; 2 - $n_{H_2} = 10^5$~cm$^{-3}$,
$X=10^6$, $T_{kin}=20$~K, and $T_{bg}=70~K$
\vskip 5mm
\renewcommand{\arraystretch}{1.2}
\tabcolsep=2.95mm
\begin{tabular}{l|r<{\rule{3mm}{0mm}}|r|r|r||l|r<{\rule{3mm}{0mm}}|r|r|r}
\tablecaption
$0_{0}-1_{-1}$ & 108.9 & 2.0\hphantom{*} & M2$^a$ & $-$2.0* & $4_{0}-3_{1}$ & 28.3 & -0.7* & $-$M$^g$ & 3.0\hphantom{*}\\
$1_{0}-0_{0}$ & 48.4 & $-$0.2* &  & 2.4\hphantom{*} & $4_{0}-3_{0}$ & 193.4 & 0.7\hphantom{*} &  & 7.8\hphantom{*}\\
$1_{0}-1_{-1}$ & 157.3 & 2.6\hphantom{*} &  & 1.6\hphantom{*} & $4_{0}-4_{-1}$ & 157.2 & 9.8\hphantom{*} & M2$^b$ & 0.1\hphantom{*}\\
$1_{0}-2_{-1}$ & 60.5 & 3.8\hphantom{*} &  & $-$4.3* & $4_{1}-3_{2}$ & 168.6 & 0.7\hphantom{*} &  & $-$0.2*\\
$1_{1}-0_{0}$ & 213.4 & 0.4\hphantom{*} &  & 3.0\hphantom{*} & $4_{1}-3_{1}$ & 193.5 & 0.8\hphantom{*} &  & 7.4\hphantom{*}\\
$1_{1}-1_{0}$ & 165.1 & 1.0\hphantom{*} & $-$M$^a$ & 0.4\hphantom{*} & $4_{1}-4_{0}$ & 165.2 & 3.2\hphantom{*} & $-$M$^a$ & 1.2\hphantom{*}\\
$1_{1}-2_{0}$ & 68.3 & 0.4\hphantom{*} &  & $-$1.1* & $4_{2}-3_{2}$ & 193.5 & 1.0\hphantom{*} &  & 2.5\hphantom{*}\\
$2_{-2}-3_{-1}$ & 278.3 & 1.2\hphantom{*} &  & 0.1\hphantom{*} & $4_{2}-3_{1}$ & 218.4 & $-$1.2* &  & 13.9\hphantom{*}\\
$2_{-1}-1_{-1}$ & 96.7 & $-$1.0* &  & 4.7\hphantom{*} & $4_{2}-4_{1}$ & 24.9 & $-$2.3* & M1$^e$ & 5.3\hphantom{*}\\
$2_{0}-1_{0}$ & 96.7 & $-$0.1* &  & 4.4\hphantom{*} & $4_{3}-3_{3}$ & 193.5 & 0.1\hphantom{*} &  & 1.0\hphantom{*}\\
$2_{0}-1_{-1}$ & 254.0 & 0.8\hphantom{*} &  & 1.9\hphantom{*} & $4_{3}-5_{2}$ & 288.7 & 0.9\hphantom{*} &  & 0.1\hphantom{*}\\
$2_{0}-2_{-1}$ & 157.3 & 6.2\hphantom{*} & M2$^b$ & $-$0.3* & $5_{-4}-4_{-4}$ & 241.8 & 0.0\hphantom{*} &  & 0.6\hphantom{*}\\
$2_{0}-3_{-1}$ & 12.2 & 4.9\hphantom{*} & M2$^c$ & $-$6.7* & $5_{-4}-6_{-3}$ & 234.7 & 0.0\hphantom{*} &  & 0.0*\\
$2_{1}-1_{1}$ & 96.8 & $-$0.2* &  & 3.5\hphantom{*} & $5_{-3}-4_{-3}$ & 241.9 & 0.2\hphantom{*} &  & 1.7\hphantom{*}\\
$2_{1}-1_{0}$ & 261.8 & 0.8\hphantom{*} &  & 4.4\hphantom{*} & $5_{-2}-4_{-2}$ & 241.9 & 1.3\hphantom{*} &  & 3.9\hphantom{*}\\
$2_{1}-2_{0}$ & 165.1 & 1.6\hphantom{*} & $-$M$^a$ & 1.0\hphantom{*} & $5_{-2}-6_{-1}$ & 133.6 & 2.8\hphantom{*} & $-$M$^h$ & $-$0.6*\\
$2_{1}-3_{0}$ & 20.0 & 0.6\hphantom{*} & M2$^d$ & $-$2.6* & $5_{-1}-4_{0}$ & 84.5 & $-$3.7* & M1$^i$ & 4.4\hphantom{*}\\
$2_{2}-1_{1}$ & 121.7 & $-$2.2* &  & 22.2\hphantom{*} & $5_{-1}-4_{-1}$ & 241.8 & 3.1\hphantom{*} &  & 6.3\hphantom{*}\\
$2_{2}-2_{1}$ & 24.9 & $-$1.0* & M1$^e$ & 7.6\hphantom{*} & $5_{0}-4_{1}$ & 76.5 & $-$0.7* &  & 2.6\hphantom{*}\\
$3_{-2}-2_{-2}$ & 145.1 & $-$0.3* &  & 2.6\hphantom{*} & $5_{0}-4_{0}$ & 241.7 & 1.5\hphantom{*} &  & 6.9\hphantom{*}\\
$3_{-2}-4_{-1}$ & 230.0 & 2.1\hphantom{*} & $-$M$^a$ & $-$0.1* & $5_{0}-5_{-1}$ & 157.2 & 9.6\hphantom{*} & M2$^b$ & 0.6\hphantom{*}\\
$3_{-1}-2_{-1}$ & 145.1 & 0.1\hphantom{*} &  & 6.9\hphantom{*} & $5_{1}-4_{2}$ & 216.9 & 1.1\hphantom{*} &  & 0.4\hphantom{*}\\
$3_{0}-2_{0}$ & 145.1 & 0.2\hphantom{*} &  & 7.7\hphantom{*} & $5_{1}-4_{1}$ & 241.9 & 1.4\hphantom{*} &  & 6.4\hphantom{*}\\
$3_{0}-3_{-1}$ & 157.3 & 8.7\hphantom{*} & M2$^b$ & $-$0.5* & $5_{1}-5_{0}$ & 165.4 & 3.9\hphantom{*} & $-$M$^a$ & 1.2\hphantom{*}\\
$3_{1}-2_{2}$ & 120.2 & 0.2\hphantom{*} &  & $-$0.5* & $5_{2}-4_{2}$ & 241.9 & 1.7\hphantom{*} &  & 3.1\hphantom{*}\\
$3_{1}-2_{1}$ & 145.1 & 0.2\hphantom{*} &  & 6.5\hphantom{*} & $5_{2}-4_{1}$ & 266.8 & $-$0.3* &  & 9.3\hphantom{*}\\
$3_{1}-3_{0}$ & 165.1 & 2.3\hphantom{*} & $-$M$^a$ & 0.9\hphantom{*} & $5_{2}-5_{1}$ & 25.0 & $-$2.2* & M1$^e$ & 3.0\hphantom{*}\\
$3_{2}-2_{2}$ & 145.1 & 0.0* &  & 1.6\hphantom{*} & $5_{3}-4_{3}$ & 241.8 & 0.4\hphantom{*} &  & 1.6\hphantom{*}\\
$3_{2}-2_{1}$ & 170.1 & $-$2.0* &  & 19.0\hphantom{*} & $5_{3}-6_{2}$ & 240.2 & 1.2\hphantom{*} &  & 0.1\hphantom{*}\\
$3_{2}-3_{1}$ & 24.9 & $-$2.0* & M1$^e$ & 7.9\hphantom{*} & $5_{4}-4_{4}$ & 241.8 & 0.0\hphantom{*} &  & 0.5\hphantom{*}\\
$4_{-4}-5_{-3}$ & 283.1 & 0.0\hphantom{*} &  & 0.0\hphantom{*} & $6_{-5}-5_{-5}$ & 290.1 & 0.0\hphantom{*} &  & 0.3\hphantom{*}\\
$4_{-3}-3_{-3}$ & 193.5 & 0.0\hphantom{*} &  & 1.2\hphantom{*} & $6_{-4}-5_{-4}$ & 290.2 & 0.1\hphantom{*} &  & 1.0\hphantom{*}\\
$4_{-2}-3_{-2}$ & 193.5 & 0.5\hphantom{*} &  & 3.5\hphantom{*} & $6_{-4}-7_{-3}$ & 186.3 & 0.0\hphantom{*} &  & 0.0*\\
$4_{-2}-5_{-1}$ & 181.8 & 2.7\hphantom{*} &  & $-$0.4* & $6_{-3}-5_{-3}$ & 290.2 & 0.3\hphantom{*} &  & 1.8\hphantom{*}\\
$4_{-1}-3_{0}$ & 36.2 & $-$4.5* & M1$^f$ & 5.5\hphantom{*} & $6_{-2}-5_{-2}$ & 290.3 & 1.9\hphantom{*} &  & 3.7\hphantom{*}\\
$4_{-1}-3_{-1}$ & 193.4 & 1.6\hphantom{*} &  & 6.8\hphantom{*} & $6_{-2}-7_{-1}$ & 85.6 & 2.8\hphantom{*} &   & $-$0.6*\\
\hline

\end{tabular}
}
\twocolumn[
\parbox{\linewidth}{
\hyphenation{kinetic abundances densities number}
\vskip -1mm
{\bf Table.} (Contd.)
\vskip 2mm
\renewcommand{\arraystretch}{1.09}
\tabcolsep=2.95mm
\begin{tabular}{l|r<{\rule{3mm}{0mm}}|r|r|r||l|r<{\rule{3mm}{0mm}}|r|r|r}
\tablecaption
$6_{-1}-5_{0}$ & 132.9 & $-$2.7* & M1$^g$ & 3.4\hphantom{*} & $8_{2}-8_{1}$ & 25.3 & $-$1.3* & M1$^e$ & 0.3\hphantom{*}\\
$6_{-1}-5_{-1}$ & 290.1 & 4.2\hphantom{*} &  & 5.5\hphantom{*} & $8_{3}-9_{2}$ & 94.5 & 0.8\hphantom{*} &  & $-$0.1*\\
$6_{0}-5_{1}$ & 124.6 & $-$0.6* &  & 2.2\hphantom{*} & $9_{-5}-10_{-4}$ & 268.7 & 0.0\hphantom{*} &  & 0.0\hphantom{*}\\
$6_{0}-5_{0}$ & 289.9 & 2.4\hphantom{*} &  & 5.7\hphantom{*} & $9_{-4}-10_{-3}$ & 41.1 & 0.0* &  & $-$0.1*\\
$6_{0}-6_{-1}$ & 157.0 & 8.8\hphantom{*} & M2$^b$ & 0.7\hphantom{*} & $9_{-3}-10_{-2}$ & 282.0 & 0.4\hphantom{*} &  & 0.0\hphantom{*}\\
$6_{1}-5_{2}$ & 265.3 & 1.4\hphantom{*} &  & 0.9\hphantom{*} & $9_{-1}-8_{0}$ & 278.3 & $-$0.5* &  & 1.5\hphantom{*}\\
$6_{1}-5_{1}$ & 290.3 & 1.9\hphantom{*} &  & 5.2\hphantom{*} & $9_{-1}-8_{-2}$ & 9.9 & $-$2.0* & M1$^k$ & 0.5\hphantom{*}\\
$6_{1}-6_{0}$ & 165.7 & 4.0\hphantom{*} &  & 1.0\hphantom{*} & $9_{0}-8_{1}$ & 267.4 & 0.1\hphantom{*} &  & 1.1\hphantom{*}\\
$6_{2}-5_{2}$ & 290.3 & 2.3\hphantom{*} &  & 3.3\hphantom{*} & $9_{0}-9_{-1}$ & 156.0 & 5.1\hphantom{*} &   & 0.5\hphantom{*}\\
$6_{2}-6_{1}$ & 25.0 & $-$1.9* & M1$^e$ & 1.5\hphantom{*} & $9_{1}-9_{0}$ & 167.9 & 2.4\hphantom{*} &  & 0.5\hphantom{*}\\
$6_{3}-5_{3}$ & 290.2 & 0.7\hphantom{*} &  & 1.9\hphantom{*} & $9_{2}-9_{1}$ & 25.5 & $-$1.0* & M1$^l$ & 0.1\hphantom{*}\\
$6_{3}-7_{2}$ & 191.7 & 1.2\hphantom{*} &  & 0.0* & $9_{3}-10_{2}$ & 45.8 & 0.6\hphantom{*} &  & $-$0.1*\\
$6_{4}-5_{4}$ & 290.2 & 0.1\hphantom{*} &  & 0.8\hphantom{*} & $10_{-5}-11_{-4}$ & 220.4 & 0.0\hphantom{*} &  & 0.0\hphantom{*}\\
$6_{5}-5_{5}$ & 290.1 & 0.0\hphantom{*} &  & 0.2\hphantom{*} & $10_{-3}-11_{-2}$ & 232.9 & 0.2\hphantom{*} &  & 0.0\hphantom{*}\\
$7_{-4}-8_{-3}$ & 137.9 & 0.0\hphantom{*} &  & $-$0.1* & $10_{-1}-9_{-2}$ & 57.3 & $-$1.4* &  & 0.4\hphantom{*}\\
$7_{-2}-8_{-1}$ & 37.7 & 2.4\hphantom{*} & M2$^j$ & $-$0.6* & $10_{0}-10_{-1}$ & 155.3 & 3.7\hphantom{*} &  & 0.4\hphantom{*}\\
$7_{-1}-6_{0}$ & 181.3 & $-$1.8* &  & 2.6\hphantom{*} & $10_{1}-10_{0}$ & 169.3 & 1.6\hphantom{*} &  & 0.3\hphantom{*}\\
$7_{0}-6_{1}$ & 172.4 & $-$0.3* &  & 1.8\hphantom{*} & $10_{2}-10_{1}$ & 25.9 & $-$0.7* &  & 0.0\hphantom{*}\\
$7_{0}-7_{-1}$ & 156.8 & 7.7\hphantom{*} & M2$^b$ & 0.7\hphantom{*} & $11_{-5}-12_{-4}$ & 172.1 & 0.0\hphantom{*} &  & 0.0*\\
$7_{1}-7_{0}$ & 166.2 & 3.7\hphantom{*} &  & 0.8\hphantom{*} & $11_{-3}-10_{-4}$ & 7.3 & 0.0\hphantom{*} &  & 0.1\hphantom{*}\\
$7_{2}-7_{1}$ & 25.1 & $-$1.6* & M1$^e$ & 0.7\hphantom{*} & $11_{-3}-12_{-2}$ & 183.7 & 0.1\hphantom{*} &  & 0.0\hphantom{*}\\
$7_{3}-8_{2}$ & 143.2 & 1.1\hphantom{*} &  & $-$0.1* & $11_{-1}-10_{-2}$ & 104.3 & $-$0.9* &  & 0.3\hphantom{*}\\
$8_{-4}-9_{-3}$ & 89.5 & 0.0\hphantom{*} &  & $-$0.1* & $11_{0}-11_{-1}$ & 154.4 & 2.4\hphantom{*} &  & 0.3\hphantom{*}\\
$8_{-1}-7_{0}$ & 229.8 & $-$1.0* & M1$^a$ & 2.0\hphantom{*} & $11_{1}-11_{0}$ & 171.2 & 0.8\hphantom{*} &  & 0.3\hphantom{*}\\
$8_{0}-7_{1}$ & 220.1 & $-$0.1* &  & 1.4\hphantom{*} & $11_{2}-10_{3}$ & 2.9 & $-$0.3* &  & 0.1\hphantom{*}\\
$8_{0}-8_{-1}$ & 156.5 & 6.5\hphantom{*} & M2$^b$ & 0.6\hphantom{*} & $11_{2}-11_{1}$ & 26.3 & $-$0.4* &  & 0.0*\\
$8_{1}-8_{0}$ & 166.9 & 3.2\hphantom{*} &  & 0.6\hphantom{*} & $11_{4}-12_{3}$ & 253.9 & 0.0\hphantom{*} &  & 0.0\hphantom{*}\\
\hline

\end{tabular}
\vskip 3mm
\footnotesize
Note:\hfill\parbox[t]{0.95\linewidth}{
M1~-- a class I maser was observed; M2~-- a class II maser was observed;
$-$M~-- no maser was detected; *~-- the optical depth in this transition
is negative; (a)~-- Slysh\etal (1999); (b)~-- Slysh\etal (1995); (c)~--
Batrla\etal (1987); (d)~-- Wilson\etal (1985); (e)~--\lb Barret\etal (1971);
(f)~-- Turner\etal (1972); (g)~-- Slysh\etal (1992); (h)~-- Slysh\etal
(1997); (i)~-- Zuckerman\etal (1972); (j)~--\lb Haschick\etal (1989);
(k)~-- Slysh\etal (1993); (l)~-- Menten\etal (1986a).
}
\vskip 2mm
}
]
\hyphenation{contains}
\noindent
Menten\etal (1986a). 
Inversion in class I models is also observed for the transitions of the
series (J+1)$_2-$J$_1$. The first two transitions in the series
($J=1$ and 2) turn\lb out to show the greatest inversion.
In model~II,~
inver\-sion is observed in the $3_1-2_2$ transition from~the~series
(J+1)$_1-$J$_2$. Since the series J$_{-2}-$(J+1)$_{-1}$~contains\lb
transitions from $K=-2$ ladder to the $K=-1$ ladder and vice versa, some
of the transitions show inversion\lb in model I ($9_{-1}-8_{-2}$,
$10_{-1}-9_{-2}$, etc.), and others show inversion in
model II ($5_{-2}\!-\!6_{-1}$, $6_{-2}\!-\!7_{-1}$, $7_{-2}\!-\!8_{-1}$;~
tran\-sitions
with smaller $J$ have higher frequencies and show no inversion or their
inversion is marginal). The lowest frequency transitions $7_{-2}-8_{-1}$
at 37~GHz and $9_{-1}-8_{-2}$\lb at 9.9~GHz, which were observed in the
direction of\lb star-forming regions (Haschick\etal 1989; Slysh\etal 1993),
are the most intense transitions in this series. No\lb maser emission was
detected in the $5_{-2}\!-\!6_{-1}$ and $3_{-2}\!-\!4_{-1}$\lb transitions
(Slysh\etal 1999).
The series (J+1)$_2-$J$_3$ (J$>9$) gives inversion
in model I. The optical depth in\lb transitions between ladders with
different $K$ is small\lb because of the low population (although inversion
may exist). Our computations are in qualitative agreement with the data
of Cragg\etal (1992), which were\lb obtained by the LVG method.
Candidates for class II masers coincide with those of Sobolev\etal (1997),\lb
with the exception of the series J$_1-$J$_0$ at 165~GHz. This result
may be related to the difference in the model\lb parameters. Peak optical
depth for the lines with frequencies below 300~GHz up to $J=11$
inclusive in models I and II are given in the table. When computing the\lb
spectra, we assumed that all sources were observed\lb\lb against the background
radiation with a temperature of\lb 2.7~K.

\vskip 3mm
\section{CONCLUSION}
\vskip 2mm

(1) The method of Bernes (1979) is applicable if the cloud
is divided into shells of equal volume, if the paths of model
photons and the optical depths in each step are\lb sufficiently
small, and if the column densities of the\lb emitting molecule in the
direction of propagation of the model photon are equal for each step.
The applicability\lb of the method proposed here is restricted only by
the convergence of the iterative procedure, which may be hampered
in the presence of strong masers.
\vskip 2mm

(2) The populations within the ladder (Figs.~1 and~2)
in the segments before and after the break are well\lb described by the
Boltzmann formula, while the populations between the ladders have a non-LTE
distribution. For this reason, no masers are formed in transitions that
occur with no change in the quantum number $K$. The position of the
break in the "logarithm of population-to-statistical-weight ratio versus
level energy" diagram\lb appears to be determined by the ratio of the rates
of~col\-lisions and radiative processes.
\vskip 2mm

(3) The transitions $2_2-1_1$ at 121~GHz, $3_2-2_1$ at 170~GHz,
$10_2-10_1$ and $11_2-11_1$ at 26~GHz, $5_0-4_1$ at\lb 76~GHz,
$6_0-5_1$ at 124~GHz, $7_0-6_1$ at 172~GHz, $7_{-1}-6_0$\lb at 181~GHz,
$10_{-1}-9_{-2}$ at 57~GHz, $11_{-1}-10_{-2}$ at 104.3~GHz,
$11_2-10_3$ at 2.9~GHz are candidates for\lb new class I masers,
while the transitions $1_0-2_{-1}$ at 61~GHz, $1_1-2_0$ at 68~GHz, and
$6_{-2}-7_{-1}$ at 86~GHz are\lb candidates for class II masers.

\vskip 7mm
\section{ACKNOWLEDGEMENTS}
\vskip 2mm

I wish to thank V.I.~Slysh for several valuable\lb remarks and
helpful discussions which undoubtedly\lb improved the content of
this paper, A.M.~Dzura for\lb valuable recommendations and remarks on the
implementation of Bernes' algorithm, and S.V.~Kalenskii for
valuable remarks on the pumping mechanism. I also\lb wish to thank
A.M.~Sobolev and A.A.~Kalinin for a discussion of the method. This
study was supported in part by the Russian Foundation for Basic
Research (project nos. 97-02-27241 and 96-02-00848) and the Radio\lb
\lb\lb\lb\lb\lb\lb\lb\lb
Astronomy Research and Education Center (project\lb no. 315).

\vskip 3mm
\section{REFERENCES}
\vskip 2mm

\noindent
Barrett,\,A.H., Schwartz,\,P.R., and Waters,\,J.W.,
{\it Astrophys.\,J.}, 1971, vol. 168, p. L101.\\
\noindent
Batrla,\,W., Mattheus,\,H.E., Menten,\,K.M., and
Walmsley\,C.M., {\it Nature}, 1987, vol. 326, p. 49.\\
\noindent
Bernes,\,C.,{\it Astron. Astrophys.}, 1979, vol. 73, p. 67. \\
\noindent
Cragg,\,D.M., Johns,\,K.P., Godfrey,\,P.D., Brown, R.D.,
{\it Mon. Not. R. Astron. Soc.},\,1992, vol.\,259, p. 203.\\
\noindent
Haschick,\,A.D., Baan,\,W.A., and Menten,\,K.M.,
{\it Astrophys. J.}, 1989, vol. 346, p. 330.\\
\noindent
Juvela,\,M., {\it Astron. Astrophys.}, 1997, vol. 322, p. 943.\\
\noindent
Lees,\,H., {\it Can. J. Phys.}, 1974, vol. 52, p. 2250.\\
\noindent
Menten,\,K.M., {\it Proc.\,Third Haystack 
Observatory~Meeting}, Haschick,\,A.D. and Ho,\,P.T.P., Eds., 1991, p. 119.\\
\noindent
Menten,\,K.M., Reid,\,M.J., Moran,\,J.M., Wilson,\,T.L., 
Johnston,\,K.J., and Batrla,\,W., {\it Astrophys. J.}, 1988, vol. 333, p. L83.\\
\noindent
Menten,\,K.M., Walmsley,\,C.M., Henkel,\,C., and
Wilson,\,T.L.,{\it Astron. Astrophys.}, 1986a, vol. 157, p. 318.\\
\noindent
Menten,\,K.M., Walmsley,\,C.M., and Henkel,\,C., 
{\it Astron. Astrophys.}, 1986b, vol. 169, p. 271.\\
\noindent
Pickett,\,H.M., Cohen,\,E.A., Brinza,\,D.E., and
Shaefer,\,M.M., {\it J. Mol. Spectrosc.}, 1981, vol. 89, p. 542.\\
\noindent
Slysh,\,V.I., Kalenskii,\,S.V., and Val'tts,\,I.E.,
{\it Astron. Astrophys.}, 1993, vol. 413, p. L133.\\
\noindent
Slysh,\,V.I., Kalenskii,\,S.V., and Val'tts,\,I.E.,
{\it Astrophys. J.}, 1992, vol. 397, p. L43.\\
\noindent
Slysh,\,V.I., Kalenskii,\,S.V., and Val'tts,\,I.E.,
{\it Astrophys. J.}, 1995, vol. 442, p. 668.\\
\noindent
Slysh,\,V.I., Kalenskii,\,S.V., Val'tts,\,I.E., and
Golubev, V.V.,{\it Astrophys. J.}, 1997, vol. 478, p. L37,
astro-ph/9701075\\
\noindent
Slysh,\,V.I., Val'tts,\,I.E., and Kalenskii,\,S.V., 
{\it Astron. Astrophys.}, 1999 (in press)\\
\noindent
Sobolev,\,A.M., Cragg,\,D.M., and
Godfrey,\,P.D.,{\it Mon. Not. R. Astron. Soc.}, 1997, vol. 288, p. L39.\\
\noindent
Turner,\,B.E., Gordon,\,M.A., and Wrixon,\,G.T.,
{\it Astrophys. J.}, 1972, vol. 177, p. 609.\\
\noindent
Wilson,\,T.L., Walmsley,\,C.M., Menten,\,K.M., and
Hermsen,\,W.,{\it Astron. Astrophys.}, 1985, vol. 147, p. L19.\\
\noindent
Zuckerman,\,B., Turner,\,B.E., Johnson,\,D.R.,
Palmer,\,P., and Morris,\,M., {\it Astrophys. J.}, 1972, vol. 177, p. 609.\\

\hfill {\it Translated by V.Astakhov}

\end{document}